\documentstyle[proceedings,namedreferences,epsf,color,bookmath,myslides]{crckapb}
\begin{opening}
\title{
Broken Symmetry and Nonequilibrium Superfluid $^3$He}
\subtitle{\hspace*{5mm}Circular Birefringence of Propagating Transverse Currents}
\runningtitle{Broken Symmetry and Nonequilibrium Superfluid $^3$He}
\author{J.~A.~Sauls}
\institute{Northwestern University, Evanston, Illinois 60208, USA}
\end{opening}
\begin{document}
\vspace{-80 mm}\noindent{\tiny Lecture Notes for the 1999 Les Houches 
                       Winter School on \\ {\sl Topological Defects and 
                       Non-Equilibrium Dynamics of Symmetry Breaking
                       Phase Transitions}}
\vspace*{75 mm} 

\begin{abstract}
The superfluid phases of \he~provide a unique physical system to
study the dynamical effects of spontaneous symmetry breaking 
in condensed matter. The theory of superfluid \he~is grounded in
two of the most successful theories of strongly interacting
matter, Landau's Fermi-liquid theory and the BCS pairing theory
of superconductivity. These two theories were placed into a 
common theoretical framework in the late 60's. I discuss 
applications of this theory to nonequilibrium dynamics of
superfluid \he. In 1957 Landau predicted that liquid \he~would 
support propagating shear waves at low temperatures, i.e. a
{\it transverse sound} mode. Such waves have recently been
observed at low temperatures in the superfluid B-phase of liquid \he.
These observations provide a beautiful example of the effect of spontaneous 
symmetry breaking in condensed matter. I discuss the theory of
transverse wave propagation in \he~and the recent detection of these waves
by magneto-acoustic rotation of the polarization in a magnetic field.
\end{abstract}

\section{Introduction}\label{intro}

The physics of superfluid \he~is sufficiently
rich that it can provide interesting analogues of
theoretical models in astrophysics,
high-energy physics and cosmology \cite{vol99},
and more recently in the field of ultra-low-temperature
atomic gases \cite{ho96}. But the significance of
liquid \he~to theoretical physics may well be that it
provides the model system for developing and 
extending one of the most successful theories of strongly
interacting matter - {\sl the Fermi-liquid theory of
superconductivity}.\footnote{For a detailed description see \cite{ser83}.}
These lecture notes provide an introduction to non-equilibrium
superfluid \he, with applications to high-frequency excitations, including
aspects of symmetry breaking in \he~and its effects on collective
mode dynamics.

The density and current modes of liquid \he~are governed by conservation
laws for mass (or particle number) and momentum,
\be\label{Conservation_Laws}
\pder{n}{t}+\dive{\vJ}=0
\quad\,,\quad
\pder{J_i}{t}+\frac{1}{m}\pder{\Pi_{ij}}{x_j}=0
\,,
\ee
where $n(\vR,t)$ is the particle density, $\vJ(\vR,t)$ is
the particle current density, $m$ is the atomic mass of
\he~and $\Pi_{ij}(\vR,t)$ is the momentum stress tensor of the liquid.
The conservation laws are supplemented by equations relating
the density, current and stress tensor. In the hydrodynamic limit
these constitutive equations are obtained from local thermodynamics.
The assumption of hydrodynamics is that the characteristic
timescale for a disturbance from equilibrium is long compared with
the timescale for the restoration of equilibrium locally on the
scale of a typical wavelength of the disturbance.
The equations of hydrodynamics provide an accurate description
of the long wavelength, low frequency ($\omega\ll 1/\tau$) 
dynamics of \he. However, this description breaks down 
at sufficiently high frequencies or low temperatures.
The regime of applicability of hydrodynamics to liquid \he~at low
temperatures is severely restricted by Fermi statistics. The time-scale
for the restoration of local equilibrium increases rapidly below
$100\,\mbox{mK}$, $\tau(T)\simeq 1\,\mu\mbox{sec-mK}^2/T^2$.
Thus, for excitation frequencies of order $\omega/2\pi\simeq 20\,\mbox{MHz}$
the hydrodynamic regime is restricted to temperatures above
$T\simeq 15-20\,\mbox{mK}$. The physics of \he~at lower temperatures,
or higher frequencies, departs radically from the predictions of hydrodynamics.
Liquid \he~at these excitation energies is governed by coupled dynamical
equations for quasiparticle (distribution function) and Cooper pair 
(pair correlation amplitude) excitations.

\vspace*{12pt}\noindent{\large\it Transport Theory}\vspace*{12pt}

In his seminal papers on Fermi liquid theory \cite{lan56,lan57}
Landau explained how a system of strongly interacting Fermions could
exhibit both a spectrum of low-lying Fermionic excitations with a
well-defined Fermi surface, as well as low lying Bosonic modes associated
with deformations of the Fermi surface. The stability of the Fermi surface 
requires that interactions between the Fermionic excitations 
(``quasiparticles'') act to {\sl restore} the Fermi surface
to its equilibrium configuration. Under these conditions the 
``Fermi sea'' behaves as an elastic medium in which restoring
forces lead to natural oscillations, or free vibrations of the Fermi
surface, called ``zero sound modes''. The dynamical variables describing
these modes are related to the {\sl deformation} of the Fermi surface,
which is defined in terms of the deviation of the quasiparticle distribution
function from its equilibrium form,
\be
\phi(\hat{\vp},\vR;t)=
     \int d\varepsilon\,\Phi(\hat{\vp},\vR;\varepsilon,t)
\,,
\ee
where $\Phi(\hat{\vp},\vR;\varepsilon,t)$ is the distribution 
function for quasiparticles with momentum near the Fermi surface,
$\vp_f=p_f\hat{\vp}$, and excitation energies, $\varepsilon$, 
that are small compared with the Fermi energy,
$|\varepsilon|\ll E_f$.\footnote{In the microscopic formulation
of Fermi liquid theory the distribution
function is defined in terms of a correlation
function in the limit, $\hbar q \ll p_f$, $\hbar\omega\ll E_f$ \cite{ser83}.}
BCS pairing in \he~is an instability of the Fermi surface; thus, one should expect
that a description of collisionless modes in the superfluid phases
requires new dynamical variables describing the condensed phase, which
are coupled to the dynamics of the distribution function.

The transport equation for the distribution function is the central equation
of Landau's theory of normal \he,
\be\label{Landau_Boltzmann}
\left(\pder{}{t} + \vv_f\cdot\gradR\right)
\Phi(\hat{\vp},\vR;\varepsilon,t)
-
\left(\pder{\Phi_0}{\varepsilon}\right)
\,
\pder{}{t}\cU_{\mbox{\tiny tot}}(\hat{\vp},\vR;t)
= I[\Phi]
\,.
\ee
The first two terms describe the ballistic propagation of quasiparticles
with a group velocity given by the Fermi velocity, $\vv_f=v_f\hat{\vp}$.
The third term represents the action of external and internal forces
acting on the quasiparticles;
$\cU_{\mbox{\tiny tot}}(\hat\vp,\vR;t)=u_{\mbox{\tiny ext}}(\vR;t)
+\cE(\hat\vp,\vR;t)$ is the sum of the external potential
energy ($u_{\mbox{\tiny ext}}$) and the interaction energy ($\cE$) of
a quasiparticle with momentum $\vp$ on the Fermi surface with
the distribution of other non-equilibrium quasiparticles.
The derivative of the equilibrium distribution, 
$\partial \Phi_0(\varepsilon)/\partial\varepsilon$,
restricts the dynamics to low-energy excitations near the
Fermi level. Both external and internal forces accelerate the 
quasiparticles leading to smooth changes in the distribution 
function in space and time. The right side of the
transport equation determines the rate of change
of the distribution function from quasiparticle 
collisions. This term leads to irreversibility and relaxation of
the distribution function on the timescale of the mean time between 
collisions, i.e. $I[\Phi]\sim-1/\tau\,\delta\Phi$.
The collision terms are generally small for the typical
frequencies of interest, i.e. $\omega\gg 1/\tau$.
Thus, we can often neglect collisional effects, except for 
collisional broadening of otherwise sharp collective modes.

The distribution function determines the density and momentum
fluctuations associated with nonequilibrium states of \he,
\ber
n(\vR,t) = \frac{p_f^3}{3\pi^2\hbar^3} +
\frac{N_f}{1+F^s_0}\int\frac{d\Omega_{\vp}}{4\pi}
\int d\varepsilon\,\Phi(\hat{\vp},\vR;\varepsilon,t)
\,,
\\
\vJ = N_f\int\frac{d\Omega_{\vp}}{4\pi}\,\vv_f
\int d\varepsilon\,\Phi(\hat{\vp},\vR;\varepsilon,t)
\,,
\eer
where $N_f$ is the density of states at the Fermi level and $F^s_0$ is
the $\ell=0$ Landau interaction parameter (defined below).
The conservation laws for mass and momentum
follow from the transport equation and conservation 
of fermion number and momentum in collision processes:
$\int d\Omega_{\vp}\int d\varepsilon\, I[\Phi] = 0$ and
$\int d\Omega_{\vp}\hat{\vp}\int d\varepsilon\, I[\Phi] = 0$.
The continuity equation for $\vJ$ determines the stress tensor,
\be\label{Stress_Tensor_Fermi_Liquid}
\Pi_{ij} =
N_f\int\,\frac{d\Omega_{\vp}}{4\pi}\,[\vv_f]_i\,[\vp_f]_j
\int d\varepsilon\,\Phi(\hat{\vp},\vR;\varepsilon,t)
\,.
\ee
The result for $\Pi_{ij}$
follows from the microscopic theory of a Fermi liquid and is
valid above and below the superfliud transition, as well as
in the nonlinear response regime, so long as 
$u_{\mbox{\tiny ext}}\ll E_f$ \cite{mck90}. 

Normal \he~is isotropic so the eigenmodes of the Fermi surface
are given by the amplitudes, $\phi_{\ell,m}$, of the spherical
harmonic expansion of the distribution function. For example,
$\phi_{0,0}$ is the amplitude for an isotropic expansion or 
contraction of the Fermi surface, and determines the density 
fluctuation,
\be\label{n-phi00}
\delta n(\vR,t)=\frac{N_f}{1+F^s_0}\,\phi_{0,0}
\,,
\ee
while the $\ell=1$ modes determine the current density,
\be\label{J-phi1m}
\vJ(\vR,t)=\onethird N_fv_f\sum_{m=0,\pm 1}\phi_{1,m}\,\ve^{(m)}
\,,
\ee
where $\ve^{(m)}$ are unit vectors defining the three linearly 
independent current modes: the longitudinal current,
$\vJ_{l}\sim\ve^{(0)}=\hat{\vq}$, and the two circularly
polarized transverse modes, $\vJ_{\pm}\sim\ve^{(\pm)}$. The
circularly polarized basis vectors are defined by, $\ve^{(0)}=\ve_3$
and $\ve^{(\pm)}=(\ve_1\pm i\ve_2)/\sqrt{\mbox{\small $2$}}$,
where $\{\ve_1,\ve_2,\ve_3\}$ is a Cartesian triad with 
$\ve_3=\hat{\vq}$. The $\ell=2$ amplitudes are related to the five 
traceless and symmetric components of the stress tensor,
\be\label{pi-phi2m}
\pi_{ij}=\Pi_{ij}-\onethird\mbox{Tr}[\vPi]\delta_{ij}=
\twofifteenths N_fp_fv_f\sum_{m=0,\pm 1,\pm 2}\phi_{2,m}\,\vt_{ij}^{(2,m)}
\,,
\ee
where $\vt_{ij}^{(2,m)}$ are the $\ell=2$ spherical tensors
with the quantization axis chosen as $\ve^{(0)}=\hat{\vq}$ (see 
Eqs. (\ref{Spherical_Tensors}) below).
The stress tensor plays a central role in the nonequilibrium 
response of liquid \he. The solution of the transport equation
leads to constitutive equations relating the stress
tensor, the density and current response, and in the superfluid phase
to additional variables representing the dynamics of the condensate. 

In the collisionless limit restoring forces for 
spin-independent deformations of the Fermi surface originate from
\be
\cE(\hat{\vp},\vR;t)=
\int\frac{d\Omega_{\vp'}}{4\pi}\,A^s(\hat{\vp},\hat{\vp}')
\,\phi(\hat{\vp}',\vR;t)
\,.
\ee
This term represents the interaction of a quasiparticle
of momentum $p_f\hat{\vp}$ with the distribution of
quasiparticles that represent the deformation of the
Fermi surface. The function $A^s(\hat{\vp}\cdot\hat{\vp}')$ 
is the spin-independent amplitude for forward
scattering of two quasiparticles with momenta $\vp$ and $\vp'$ 
on the Fermi surface, and is related to the spin-independent 
interaction energy between two quasiparticles, 
$F^s(\hat{\vp},\hat{\vp}')$, by \cite{lan57}
\be
A^s(\hat{\vp}\cdot\hat{\vp}')=F^s(\hat{\vp}\cdot\hat{\vp}')+
\int\frac{d\Omega_{\vp''}}{4\pi}\,
F^s(\hat{\vp}\cdot\hat{\vp}'')A^s(\hat{\vp}''\cdot\hat{\vp}')
\,.
\ee

The underlying force between \he~atoms
is rotationally invariant. Thus, $A^s$ is a function only of
the relative orientation of $\vp$ and $\vp'$,
which allows us to parametrize $F^s$ (or $A^s$) by a set
of interactions for quasiparticles in relative angular momentum channels,
i.e. $F^s=\sum_{\ell}\,F^s_{\ell}\,P_{\ell}(\hat{\vp}\cdot\hat{\vp}')$,
where the $F^s_{\ell}$ are the dimensionless Landau parameters. These
interaction parameters determine many of the physical properties of the
Fermi liquid, e.g. the change in the interaction energy induced by
an {\sl isotropic deformation} of the Fermi surface, a dilatation,
is determined by the $\ell = 0$ interaction, $F^s_0$. Similarly, the
interaction parameter in the $\ell=1$ channel represents the current-current
interaction between quasiparticles. Galilean invariance of the interactions
in liquid \he~provides a relation connecting the effective mass of a
quasiparticle to the $\ell=1$ interaction, $m^*/m=(1+F^s_1/3)$.
Since the effective mass can be obtained from the low-temperature
heat capacity we obtain from fundamental measurements the 
quasiparticle interaction in the current-current channel.
Both interactions determine the compressibility and hydrodynamic
sound velocity of \he. Strong repulsive interactions in liquid \he~lead
to a sound velocity that is significantly larger than the velocity 
of quasiparticle excitations,
$c_1/v_f=\sqrt{\tinyonethird(1+F^s_0)(1+\tinyonethird F^s_1)}$.
Measurements of the heat capacity and sound velocity yield values of 
$F^s_{0}\simeq 10$ ($F^s_{1}\simeq 6$) at $p=1\,\mbox{bar}$ to 
$F^s_{0}\simeq 100$ ($F^s_{1}\simeq 15$) at $p=34\,\mbox{atm}$, corresponding
to interaction energies $\sim 10-100$ times the Fermi energy.
\footnote{The {\sl spin-dependent} interaction energy originates from
exchange interactions which are described by the spin-dependent
Landau parameters, $\{F^a_{\ell}\}$. These interactions
are important for \he~in a magnetic field; they produce exchange enhancement of
an applied magnetic field.}
Thus, although quasiparticles are long-lived single-particle
excitations at low energies and low temperatures, liquid 
\he~can never be described as a ``gas'' of weakly 
interacting quasiparticles.

An important consequence of strong interactions is that 
zero sound modes exist at low temperatures in the
collisionless regime, $\omega\tau(T)\gg 1$.
The restoring forces are provided by strong repulsive
quasiparticle interactions. The first observation of longitudinal zero
sound was reported in Ref. \cite{abe66}, and has been studied extensively
\cite{hal90}. The signatures of the cross-over from hydrodynamic to collisionless 
sound are a small change in sound velocity in the collisionless regime,
$(c_0-c_1)/c_1\simeq \tinytwofifths(1+\tinyonefifth F^s_2)/(1+F^s_0)\sim 10^{-2}$, 
and a dramatic change in the temperature dependence of the attenuation,
$\alpha_1\sim \omega^2/T^2$ for $\omega\tau <1$ compared with 
$\alpha_0\sim T^2$ for $\omega\tau >1$ \cite{abe66}.

In addition to longitudinal zero sound, Landau predicted a transverse
mode of the Fermi surface in the collisionless regime \cite{lan57}.
This mode should be observable as a propagating shear wave, or {\sl transverse
zero sound} (TZS). The velocity of TZS in normal \he~is expected to
be close to the Fermi
velocity, $c_t\simeq v_f$ \cite{fom68}, and the collisional damping is predicted to be
serveral orders of magnitude larger than the damping of the longitudinal
zero sound mode \cite{cor69}. In contrast to the detection of longitudinal
zero sound, the search for TZS has been more difficult.
Early experimental efforts to observe this mode were inconclusive 
\cite{roa76,flo76}. The recent observation of a propagating transverse current mode
in superfluid \heb~\cite{lee99} provides new insight into the dynamics of normal
and superfluid \he, as well as new information on many-body correlation
effects. Before discussing the theory of TZS and order parameter dynamics,
I review some basic facts about superfluid \he.

\section{Symmetry Breaking and the Order Parameter}\label{symmetry}

The Hamiltonian that determines the
properties of \he~has a high degree of symmetry. The
ground state electronic configuration of the \he~atom 
is a filled $1s$ shell, so the atom is isotropic and the 
interaction energy between
two \he~atoms is dominated by a central potential, $v(r)$,
which is also isotropic and spin-independent. This interaction
energy is of order $v(\bar{r})\simeq -10\,\mbox{K}$
at the minimum of the \he-\he~potential, 
and has a hard repulsive core at
$\bar{r}\approx 2.5\,\mbox{\AA}$. By comparison,
the nuclear dipolar energy at these densities is of order
$v_{\mbox{\tiny d}}=(\gamma\hbar)^2/\bar{r}^3\simeq
10^{-7}\,\mbox{K}$. The dipolar energy is tiny even in comparison to the
energy scale of the superfluid transition, 
$T_c\simeq 10^{-3}\,\mbox{K}$.
Thus, to an excellent approximation the equilibrium phase of
\he~above the superfluid transition is {\sl separately}
invariant under rotations in orbital space and in spin space.

Above $T_c$ \he~also posseses discrete symmetries under 
space-inversion ($\cP$), time-inversion ($\cT$),
and an approximate symmetry under particle to hole
conversion ($\cC$), \ie the transformation of a quasiparticle
with energy $\varepsilon=\xi_{\vp}$ ($>0$) and spin projection 
$\uparrow$ into a quasihole with the same excitation energy
($|\xi_{\underline{\vp}}|=\xi_{\vp}$) and spin projection 
$\downarrow$. The transformation is represented by a unitary operator,
$\cC^{\dag} a^{\dag}_{\vp\alpha}\cC=\left[i\sigma_y\right]_{\alpha\beta}
a_{\underline{\vp}\beta}$, where $a^{\dag}_{\vp\alpha}$ 
($a_{\underline{\vp}\beta}$) creates a quasiparticle (quasihole) 
with momentum $\vp$ and spin 
projection $\alpha$.\footnote{The quasiparticle operators are defined
on a restricted Hilbert space of states in the low-energy band
$|\xi_{\vp}|<\epsilon_c$ about the Fermi surface,
and are constructed to generate the low-energy 
quasiparticle Green's function \cite{noz64,ser83b}.}
Under the particle-hole transformation the
low-energy effective Hamiltonian for quasiparticles
is approximately invariant. Particle-hole symmetry in a Fermi
liquid is analogous to charge conjugation. The order
of magnitude of the particle-hole {\sl asymmetry} is given by 
$\zeta=N'(E_f)\varepsilon/N(E_f)\sim\left(\varepsilon/E_f\right)$.
Particle-hole symmetry becomes exact for $\varepsilon\rightarrow 0$;
at $T\simeq T_c$, particle-hole asymmetry is $\zeta\lsim 10^{-2}$.

The normal state of liquid \he~is also invariant 
under global gauge transformations,
$\psi(\vr)\rightarrow\psi'(\vr)=\psi(\vr)e^{i\chi}$.
When combined with the 
discrete symmetries under $\cP$, $\cT$ and $\cC$, 
the symmetry group of liquid \he~below 
$T\sim 100\,\mbox{mK}$ and above $T_c\sim 1\,\mbox{mK}$ is
\be
\cG = SO(3)_{\mbox{\tiny S}}\times SO(3)_{\mbox{\tiny L}}\times
U(1)_{\tiny \chi}\times \cP \times \cT \times \cC
\,.
\ee
This high degree of symmetry of liquid \he~is spontaneously 
broken at the superfluid transition. The {\sl residual symmetry} of the 
superfluid phase(s) is reflected in the invariant subgroup of 
$\cG$ defined by the order
parameter; the latter is a particular realization
of an orbital p-wave, spin-triplet amplitude
for Cooper pairs. The order parameter is
defined in terms of a superposition of quasiparticle pair states
with zero total momentum in the low-energy shell 
($|\xi_{\vp}|\le\epsilon_c\ll E_f$) about the Fermi surface,
\be\label{Time-Dependent_Gap_Equation}
\Delta_{\alpha\beta}(\hat{\vp},\vR;t)=
\int\frac{d\Omega_{\hat{\vp}'}}{4\pi}\,
V_t(\hat{\vp}\cdot\hat{\vp}')\,
\int_{-\epsilon_c}^{\epsilon_c}\,\frac{d\varepsilon}{2\pi i}\,
f_{\alpha\beta}(\hat\vp',\vR;\varepsilon,t)
\,,
\ee
where $f_{\alpha\beta}\sim\left<a_{\vp\alpha}a^\dag_{-\vp\beta}\right>$
is the amplitude for Cooper pairs in the spin state $\ket{\alpha\beta}$,
and $V_t=3V_1\hat\vp\cdot\hat\vp'$ is the pairing interaction in the 
spin-triplet, p-wave channel.
The matrix structure in spin-space represents the three 
spin-triplet amplitudes, each of which is a function of the 
orbital momentum of the pairs. The order parameter can 
be represented as \cite{bal63}
\be
\Delta_{\alpha\beta}(\hat{\vp},\vR;t)=\sum_{\mu}\sum_{i}\,
\left(i\sigma_{\mu}\sigma_{y}\right)_{\alpha\beta}\,
d_{\mu i}(\vR,t)\,\hat{\vp}_i
\,,
\ee
where the $3\times 3$ complex matrix, $d_{\mu i}$~,
transforms as a vector (with respect to $\mu$) under rotations
in spin space and, independently, as a vector 
(with respect to $i$) under rotations in orbital space.

The equilibrium B-phase of \he~is identified 
\cite{leg75} with an order parameter
belonging to the Balian-Werthamer (BW) class of p-wave states
\cite{bal63}. The simplest
BW state, $d_{\mu i}\sim\delta_{\mu i}$, corresponds to
pairing with quantum numbers $S=1$, $L=1$ and $J=0$,
where $\vJ=\vL + \vS$ is the total angular momentum.
The $J=0$ state belongs to a continuous manifold of
degenerate ground states; \ie any order parameter obtained
from the $J=0$ state by a \underline{relative} rotation
of the spin and orbital coordinates and a uniform gauge
transformation minimizes the free energy. Thus, a
general BW state is,
\be
d_{\mu i}=\frac{\Delta}{\sqrt{3}}\,e^{i\chi}\,
\cR_{\mu i}[\hat{\vn},\vartheta]
\,,
\ee
where $\cR[\hat{\vn},\vartheta]$ is an orthogonal 
matrix representing a relative rotation of the spin and
orbital coordinates about the axis $\hat{\vn}$ by the angle
$\vartheta$. Note that the
{\sl relative spin-orbit symmetry} is spontaneously
broken below $T_c$.\footnote{The degeneracy of the homogeneous BW
state is partially resolved by weak perturbations, e.g. the
nuclear dipolar energy or the magnetic Zeeman energy.}
Since parity and gauge symmetry are also broken below 
$T_c$ the residual symmetry group for \heb~includes the group
of combined spin and orbital rotations and time-inversion,
\be
\cG_{\mbox{\tiny resid}}=
SO(3)_{\mbox{\tiny L+S}}\times\cT\times\cC
\,,
\ee
with the generator for the rotation group given 
by $\vJ=\vL+\cR^{-1}\cdot\vS$.

The dynamics of the order parameter is governed by the time-dependent
`gap equation', Eq. (\ref{Time-Dependent_Gap_Equation}), where the 
pairing amplitude and quasiparticle distribution function obey
coupled quasiclassical transport equations. The coupling of the
quasiparticle and condensate dynamics is rooted in
particle-hole coherence, and is responsible for most of the novel
physics associated with nonequilibrium superfluidity \cite{kur90},
including the coupling between the mass current and the collective
modes of the order parameter. The rotational symmetry of the ground 
state of \heb~allows us to classify the order parameter 
excitations, which are Bosonic modes, in terms of the eigenvalues 
of $J^2$ and $J_3$ \cite{mak74}. Furthermore, there is a doubling
of the spectrum for each $(J,M)$ labeled by $\cC$ parity, $\zeta=\pm$.
Within the spin-triplet, p-wave subspace we can expand
the order parameter fluctuations in the
eigenfunctions of $(J^2,J_3,\cC)$,
\be
\delta d_{\mu i}=\sum_{J,M,\zeta=\pm}\,
\cD^{(\zeta)}_{J,M}\,\vt_{\mu i}^{(J,M)}
\,,
\ee
where $\vt_{\mu i}^{(J,M)}$ are spherical tensors that
transform according to the $J^{\mbox{\tiny th}}$ irreducible
representation of the rotation group,
\be\label{Spherical_Tensors}
\begin{array}{ll}
\vt_{ij}^{(0,0)}={1\over\sqrt{3}}\delta_{ij}
&
\vt_{ij}^{(2,0)}= \sqrt{3\over 2}
(\ve^{(0)}_i\ve^{(0)}_j-{1\over 3}\delta_{ij})
\\
\vt_{ij}^{(1,M)}={1\over\sqrt{2}}\varepsilon_{ijk}\,\ve^{(M)}_{k}
&
\vt_{ij}^{(2,\pm 1)}={1\over\sqrt{2}}(\ve^{(0)}_i\ve_j^{(\pm)}
+\ve_{i}^{(\pm)}\ve^{(0)}_j)
\\
&
\vt_{ij}^{(2,\pm 2)}=\ve_i^{(\pm)}\ve_j^{(\pm)}
\end{array}
\,.
\ee
The tensors are defined in terms of the circularly
polarized basis vectors, $\ve^{(0)}=\ve_3$ and
$\ve^{(\pm)}=(\ve_1\pm i\ve_2)/\sqrt{\mbox{\small $2$}}$,
where $\{\ve_1,\ve_2,\ve_3\}$ is a Cartesian triad of unit
vectors defining the coordinate system for the Cooper pairs.

\begin{table}\label{table-modes}
\begin{center}
\begin{tabular}{|l|l|l|c|cl|}
\hline
$\cD^{(\zeta)}_{J,M}$ & Mode & Frequency & Degeneracy &
                                         \hspace*{2 em}Coupling to & $\vPi$ \\
\hline
\hline
${\cD}^{({-})}_{0,0}$ & phase & $\omega=
\frac{1}{\sqrt{3}} v_f q$ & 1 & 1 & (L) \\
\hline
${\cD}^{({+})}_{0,0}$ & amplitude & $\omega=2\Delta$ & 1 &
                          $\zeta\left(\frac{qv_f}{\Delta}\right)$ & (L)\\
\hline
${\cD}^{({-})}_{1,M}$ & amplitude & $\omega=2\Delta$ & 3 & - & - \\
\hline
${\cD}^{({+})}_{1,M}$ & spin-waves &
$\omega=\frac{1}{\sqrt{3}} v_f q$ & 3 & -  & - \\
\hline
{${\cD}^{({-})}_{2,M}$} & {{\Im squashing}} &{
{$\omega=\sqrt{\frac{12}{5}}\Delta$}} & {{5}} &
                          $\left(\frac{\omega}{\Delta}\right)$ & (L,T)\\
\hline
${\cD}^{({+})}_{2,M}$ & \Re squashing &
$\omega=\sqrt{\frac{8}{5}}\Delta$ & 5 &
                          $\zeta\left(\frac{qv_f}{\Delta}\right)$ & (L,T)\\
\hline
\end{tabular}
\caption{The order parameter collective mode spectrum for p-wave,
spin-triplet pairing fluctuations in \heb. There are four
Goldstone modes associated with the spontaneous breaking of
gauge and relative spin-orbit rotational symmetry.
All other modes are `massive' and
correspond to fluctuations of the order parameter that are not
related by a long wavelength rotation or gauge transformation
of the ground state order parameter.}
\end{center}
\end{table}

There are $2$ (gauge) $\times$ $3$ (spin) $\times$ $3$ (orbital) $=18$
p-wave, spin-triplet Bosonic modes of \heb, which are listed in 
Table \ref{table-modes} and labeled by the quantum numbers, 
$(J^{\zeta},M)$. The eigenfrequencies are given for 
the limit $q\rightarrow 0$ {\sl neglecting interactions}. 
There are four Goldstone modes associated with the spontaneous 
breaking of gauge and relative spin-orbit rotational symmetry.
These modes are observable
because they couple to the mass and spin currents.
The $J=0^-$ mode is the phase mode \cite{and58,bog58}, 
which is essential for understanding the propagation of 
{\sl longitudinal sound} in superfluid \he.
The Goldstone modes with $J=1^{+}$ correspond to long 
wavelength fluctuations in the axis of rotation, 
$\vn$, and the angle of rotation, $\vartheta$, and are
related to the NMR and spin dynamics in \heb~\cite{leg75}.
All other modes have an excitation energy of order $\Delta$
and correspond to deformations of the order parameter that are unrelated by 
a rotation or gauge transformation to the ground state order
parameter.

\begin{figure}[ht]
\centerline{\epsfxsize0.9\hsize
\epsfbox{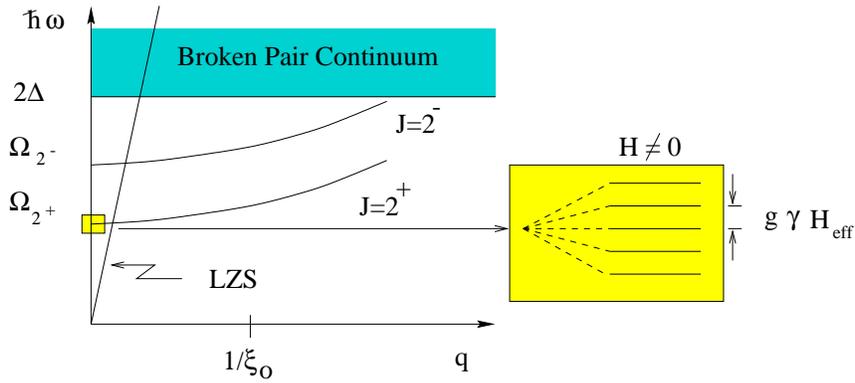}}
\caption{\small\label{Mode-Spectrum}
Dispersion relations for the uncoupled acoustic and $J=2^{\pm}$
order parameter collective modes of \heb. Longitudinal zero sound (LSZ) has
a steep dispersion relation that crosses the $J=2^{\pm}$
modes that lie below the pairbreaking continuum
at $2\Delta$. The nuclear Zeeman splitting of the $J=2^+$ modes
is also shown. The level splitting is determined by the effective Larmor
frequency, $\omega_{\mbox{\tiny eff}}=\gamma H_{\mbox{\tiny eff}}$,
and is small compared with the mode frequency for fields 
below $H\simeq 1\,\mbox{kG}$.
}
\end{figure}

Figure \ref{Mode-Spectrum} shows the spectrum of
uncoupled acoustic and $J=2$ order parameter collective modes 
of superfluid \heb.  The pairbreaking continuum onsets at $\omega=2\Delta(T)$.
Below the pairbreaking edge the zero sound mode crosses the dispersion curves for
the $J=2^{\pm}$ order parameter collective modes. Resonant absorption and
anomalous dispersion of longitudinal sound result from the coupling of these
modes to density, current and stress fluctuations \cite{wol73,mak74}.
Quasiparticle interactions lead to renomalization of the mode frequencies
\cite{sau81a}, as well as phenomena that cannot be anticipated from
a theory based on uncoupled bosonic modes and weakly interacting
quasiparticles \cite{mck90}.

Sound propagates in normal liquid \he~because quasiparticle
interactions and collisions provide restoring forces against
density fluctuations.
The restoring forces enter through the stress tensor.
In the collisionless regime 
the main contributions to the induced stress fluctuations in
normal \he~come from the interaction of a quasiparticle with the density
and current fluctuations. Both channels
contribute to the longitudinal zero sound velocity; however, only
the current fluctuations contribute to the restoring force for transverse
zero sound. In the superfluid phases new physics enters because
dynamical order parameter fluctuations also contribute to the stress tensor,
and therefore couple to density and current fluctuations.

The general form of the stress tensor follows from symmetry
considerations. In \heb~the stress fluctuations are described by
a symmetric second-rank tensor under the group of joint rotations,
$SO(3)_{\mbox{\tiny L+S}}$,
\ber
\hspace*{1cm}
\delta\Pi_{ij}
&&
         \simeq\, \Pi_n\,\delta n\,\delta_{ij}
         + \,\Pi_J\left(J_i q_j+J_j q_i\right)
\nonumber
\\
&&       + \,\Pi_{d+}\left(\delta d^{(+)}_{ij} + \delta d^{(+)}_{ji}\right)
         + \,\Pi_{d-}\left(\delta d^{(-)}_{ij} + \delta d^{(-)}_{ji}\right)
\,,
\eer
where $\delta d_{ij}^{(\pm)}$ represents order parameter fluctuations
which are even $(+)$ and odd $(-)$ under particle-hole conversion, i.e.
$\delta d_{ij}^{(\pm)}{\buildrel\cC\over\rightarrow}\pm\delta d_{ij}^{(\pm)}$. 
In the limit of exact particle-hole symmetry 
the density, current and stress fluctuations are odd under $\cC$:
$\delta n\rightarrow-\delta n$, $\vJ\rightarrow -\vJ$ and 
$\delta\Pi_{ij}\rightarrow-\delta\Pi_{ij}$.
This implies that coupling of the stress to the order parameter 
fluctuations, $\delta d_{ij}^{(+)}$, is non-vanishing only because of
particle-hole {\sl asymmetry} \cite{koc81,ser83b}. Thus, the relative 
magnitude of the $J=2^{\pm}$ coupling is $\Pi_{d+}\sim(\Delta/E_f)\Pi_{d-}$.
The contributions of these dynamical fluctuations to the stress tensor 
are obtained by solving the transport equations for the quasiparticle
distribution function and paring amplitude.

For the acoustic modes, the analysis of the transport equations is simplified
by expressing the conservation laws for particle number and momentum 
in terms of the amplitudes of the distribution function, $\phi_{\ell m}$,
using Eqs. (\ref{n-phi00})-(\ref{J-phi1m}).  The generalization of the
amplitudes, $\phi_{\ell,m}$, to the superfluid phases is discussed for example
in \cite{mck90}.

The continuity equation for the 
number density is equivalent to,
\be\label{Number_Conservation_phi}
\omega\,\phi_{0,0}-\onethird\left(1+F^s_0\right)qv_f\,\phi_{1,0} = 0
\,,
\ee
while the equations for momentum conservation can be expressed as,
\be\label{Longitudinal_Current_Conservation_phi}
\omega\,\phi_{1,0}-\left(1+\onethird F^s_1\right)\,
qv_f\,\left(\phi_{0,0}+\twofifths\sqrt{\twothirds}\,\phi_{2,0}\right) = 0
\,,
\ee
for the longitudinal current, and
\be\label{Transverse_Current_Conservation_phi}
\omega\,\phi_{1,\pm 1}-
\left(1+\onethird F^s_1\right)\,qv_f\,
\twofifths\sqrt{\onehalf}\,\phi_{2,\pm 1} = 0
\,,
\ee
for the transverse current.

\section{Longitudinal Modes}\label{lmodes}

The dispersion relation for longitudinal sound is obtained by
combining Eqs. (\ref{Number_Conservation_phi}) and 
(\ref{Longitudinal_Current_Conservation_phi}),
\be
\omega^2 -c^2_1q^2
\left\{1+\twofifths\sqrt{\twothirds}\,\xi(\omega,q)\right\}=0
\,,
\ee
where $c_1$ is the hydrodynamic sound velocity. The effects of 
mode coupling are contained in the response function,
\be\label{longitudinal-response}
\xi(\omega,q)=
\left[\frac{\delta\phi_{2,0}}{\delta\phi_{0,0}}\right]_{\mbox{\tiny tot}} =
\left(\frac{\delta\phi_{2,0}}{\delta\phi_{1,0}}\right)
\left(\frac{\delta\phi_{1,0}}{\delta\phi_{0,0}}\right)                    + ...
\,,
\ee
which represents the total stress induced by a density fluctuation.
To calculate the response function we solve the coupled dynamical
equations for the quasiparticle distribution function, time-dependent
order parameter, and the Landau 
interactions self-consistently, {\sl c.f.} \cite{mck90}. The 
solution for the longitudinal component of the induced stress is
\ber
\phi_{2,0} 
&=& A^s_0\,\alpha_0(q,\omega)\,\phi_{0,0}
+\onethird A^s_1\,\alpha_1(q,\omega)\,\phi_{1,0}
\nonumber
\\
&+&\left(\frac{\omega}{2\Delta}\right)\,
\left[\beta_0(q,\omega)\,\cD^{(-)}_{0,0}
+     \beta_2(q,\omega)\,\cD^{(-)}_{2,0}
\right]
\nonumber
\\
&+&\left(\frac{qv_f}{2\Delta}\right)\,
\left[\zeta_0(q,\omega)\,\cD^{(+)}_{0,0}
+     \zeta_2(q,\omega)\,\cD^{(+)}_{2,0}
\right]
\,,
\eer
where $\alpha_0=\xi_0$,
$\alpha_1=(qv_f/\omega)[\sqrt{\tinytwothirds}\xi_2+\tinyonethird\xi_1]$,
$\beta_0=\tinyfifteenovertworoottwo[\lambda_1-\tinyonethird\lambda_0]$,
$\beta_2=\tinyfortyfivefourths[\lambda_2-\tinytwothirds\lambda_1+\tinyoneninth\lambda_0]$,
and the functions, $\xi_0$, $\xi_1$, $\lambda_0$, $\lambda_1$ and
$\lambda_2$ are given in Ref. \cite{moo93}. In addition to the density,
$\phi_{0,0}$, and longitudinal current fluctuation, $\phi_{1,0}$,
the order parameter modes with $J=0,2$ and $M=0$ contribute to the longitudinal 
stress.
The stress induced by the $J=0^+$ and $J=2^+$ modes is small compared with
the $J=0^-$ and $J=2^-$ modes by the particle-hole asymmetry parameter, $\zeta$.
For the particle-hole asymmetric couplings, $\zeta_{0,2}$ see \cite{koc81}.
Unless stated otherwise, I will omit the contributions to the stress tensor
coming from the Fermi-liquid interactions with $\ell\ge 2$ as well as higher
angular momentum pairing channels.

The equations for the order parameter modes,
$\cD_{0,0}^{(-)}$ and $\cD_{2,0}^{(-)}$, obtained from solutions
of the time-dependent gap equation, reduce to,
\be\label{D00-}
\left[\omega^2-\frac{1}{3}(qv_f)^2\right]\cD_{0,0}^{(-)}=
\left[2\sqrt{3}\left(\omega\Delta\right)\phi_{0,0}
+\frac{2\sqrt{2}}{15}\left(qv_f\right)^2\cD_{2,0}^{(-)}\right]
\,,
\ee
\ber\label{D20-}
\left[(\omega+i\Gamma)^2- \Omega_{2^-}^2 -
       \frac{7}{5}q^2v_f^2\right]\cD_{2,0}^{(-)} &=&
\frac{2}{15}(qv_f)^2\cD_{0,0}^{(-)}
\nonumber
\\
&+&\frac{A^s_1}{3}\,
\left(\frac{qv_f}{2\Delta}\right)
\left(\frac{8}{5}\Delta^2\right)
\sqrt{\twothirds}\,\phi_{1,0}
\,.
\eer
Equation (\ref{D00-}) is the wave equation for the Anderson-Bogoliubov
phase mode, $\cD_{0,0}^{(-)}$ \cite{and58,bog58}.
For $q\rightarrow 0$ the driving term on the right side of Eq.(\ref{D00-})
is proportional to the density fluctuation, and in this limit the equation
for $\cD_{0,0}^{(-)}$ is equivalent to the Josephson equation,
$i\hbar\partial_t\chi=\delta\mu$ \cite{mck90}, 
for the order parameter phase, $\chi$, where $\delta\mu$ is the 
change in the chemical potential.
For shorter wavelengths and higher frequencies, the phase
fluctuations are also coupled to the high-frequency modes.
Equation (\ref{D20-}) describes driven oscillations of the
$J=2^-$, $M=0$ order parameter mode
by the change in the Landau interaction energy induced by
a current fluctuation; note the Landau parameter $A^s_1$.
The driving term includes the longitudinal current, $\phi_{1,0}$,
and the phase fluctuation mode, $\cD_{0,0}^{(-)}$.
The stress induced by fluctuations of the $J=2^{+}$ mode is much weaker
than that of the $J=2^{-}$ mode because of approximate particle-hole symmetry.
The $J=2^{+}$ mode is excited by density and current fluctuations, but
the coupling is again small by a factor of $\zeta$,
\be\label{D20+}
\left[(\omega+i\Gamma)^2-\Omega_{2^+}^2 -
       \frac{7}{5}q^2v_f^2\right]\cD_{2,0}^{(+)} =
2\zeta\,\omega\Delta\,\cD_{2,0}^{(-)}
\,.
\ee
Combining Eqs. (\ref{longitudinal-response})-(\ref{D20+})
we obtain the dispersion relation for longitudinal sound
\cite{wol73,koc81} with
\ber\label{xi-LZS}
\xi(q,\omega)
&=&
\frac{2}{5}\left(\frac{c_1 q}{\omega}\right)^2\,
\left(\frac{1}{1+F_0^s}\right)
\nonumber
\\
\times\Bigg\{\rho_n(\omega)
&+&
\frac{3}{5} \rho_s(\omega)\,
\left[\frac{\omega^2}{(\omega+i\Gamma)^2 - 
\Omega^2_{2^-} -\frac{7}{5}(q^2v_f^2)}\right]
\nonumber
\\
&+&
\zeta^2\,\rho_s(\omega)\,
\left[\frac{\omega^2}{(\omega+i\Gamma)^2 - 
\Omega^2_{2^+} -\frac{7}{5}(q^2v_f^2)}\right]\Bigg\}
\,.
\eer
The response function exhibits the contribution from the
quasiparticle restoring force $\sim\rho_n(\omega)$, as well as
the $J=2^{\pm}$, $M=0$, order parameter modes, $\sim\rho_s(\omega)$.
Note that $\rho_s$ and $\rho_n$ represent condensate and non-condensate
response functions with $\rho_s+\rho_n=1$; these functions reduce
to the superfluid and normal fluid densites in the 
limit $\omega=0$.\footnote{The condensate repsonse function
$\rho_s(\omega)$ is equivalent to the
Tsuneto function $\lambda(\omega)$ in \cite{moo93}.}
The contribution from the $J=2^+$ mode is also proportional to the
particle-hole asymmetry factor, $\zeta^2$. The mode couplings are
enhanced near the resonance frequency of the modes. 
The resonance linewidth is determined by quasiparticle
scattering rate which becomes exceedingly small at low temperatures,
$\Gamma(T)\simeq \Gamma_n(T_c/T)^{\tinythreehalves}\,
e^{-\Delta/T}\ll\Omega_{2^{\pm}}\sim\Delta$,
with $\Gamma_n\simeq 1/\tau(T_c)\sim 0.1\,\mbox{MHz}$ \cite{hal82}.
Thus, even weakly coupled modes are observable near resonance as strong
absorption features in the attenuation spectrum of longitudinal sound.
In addition to the resonant contributions to the absorption of
longitudinal sound, there is an absorption band for frequencies
above the pair-breaking threshold, $\omega\ge 2\Delta(T)$.
\begin{figure}[ht]
\centerline{\epsfxsize 0.75\hsize
\epsfbox{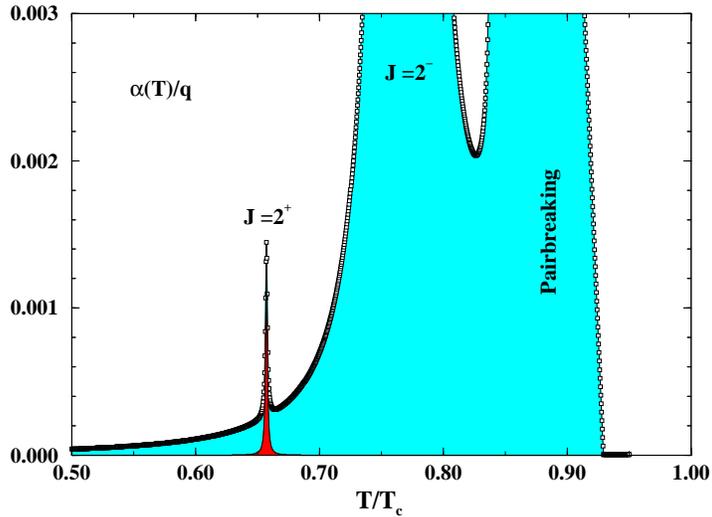}}
\caption{\small\label{longitudinal_absorption}
The calculated absorption spectrum (relative to the normal-state at $T_c$)
for longitudinal sound in \heb~at
$P\simeq 0\,\mbox{bar}$ ($T_c=0.93\,\mbox{mK}$) 
and a frequency of $35.8\,\mbox{MHz}$. The calculation
assumes $\zeta=10^{-2}$. The peaks correspond to resonant absorption 
by the $J=2^{\pm}$ modes. The relative weight of the resonance peaks 
reflects the weak coupling of the $J=2^+$ mode due to approximate
particle-hole symmetry.
}
\end{figure}
These features are shown in the calculated attenuation 
spectrum of Fig. \ref{longitudinal_absorption} as a function of
temperature for longitudinal sound at 
$P\simeq 0\,\mbox{bar}$ ($T_c=0.93\,\mbox{mK}$) and frequency of
$\omega/2\pi=35.8\,\mbox{MHz}$.
The sharp resonance peak at the lowest temperature is the
$J=2^+$ resonance, while the huge absorption peak at a
higher temperature is $J=2^-$ resonance. Pair-breaking
gives rise to the broad absorption band onsetting 
at $T_{c}$.
The $J=2^+$ mode resonance was observed in sound absorption
by \cite{gia80} and \cite{mas80}. The identification of
this resonance with the $J=2^+$ mode
was made by \cite{ave80} who observed the Zeeman splitting of the
absorption line in a magnetic field. 

\vspace*{18pt}
\noindent{\large\it Zeeman Splitting of the $J=2^{\pm}$ Modes}
\vspace*{12pt}

The $J=2$ modes are five-fold degenerate in zero field; however, 
the nuclear magnetic moment of \he~leads to a Zeeman coupling in a magnetic
field that lifts the degeneracy \cite{tew79a}, producing
five Zeeman levels in a field. For weak magnetic
fields the level splitting is linear in the field and given by
\be
\Omega_{2^{\pm},M}=\Omega_{2^{\pm}}\,+\,
                   M\,g_{2^{\pm}}\,\omega_{\mbox{\tiny eff}}
\,,
\ee
where $g_{2^{\pm}}$ is the g-factor for the $J=2^{\pm}$ modes \cite{tew79a,sau82}
and $\omega_{\mbox{\tiny eff}}$ is the effective Larmor frequency of the excited pairs,
\be\label{Larmor}
\omega_{\mbox{\tiny eff}}= \gamma H 
\frac{(1+\onefifth F_2^a)}
{1+ F^a_0(\twothirds+\onethird Y(T))+\onefifth F_2^a[\onethird+(\twothirds+F_0^a)Y(T)]}
\,,
\ee
where $Y(T)$ is the Yosida function,
and $\gamma$ is the gyromagnetic ratio for the \he~nucleus.
Equation (\ref{Larmor}) includes the exchange field enhancement of
the applied field, $H$, in the superfluid state.
The g-factors for the excited pair states, including
Fermi liquid effects and higher angular momentum pairing channels,
were calculated by \cite{sau82}.

Measurements of the Zeeman splitting
provide an experimental determination of the Lande g-factor.
The g-factor is sensitive to exchange and pairing interactions, 
and is an excellent parameter to
study many-body correlations and strong-coupling effects in
superfluid \he~\cite{sau82}. Measurements of the g-factor from 
the Zeeman splitting of the absorption spectrum of
LZS for the $J=2^+$ modes are in reasonable agreement with theoretical
calculations. However, precision measurements for the $J=2^-$ modes
have not previously been possible because the coupling of the $J=2^-$
modes is so strong that high magnetic fields are required to
resolve different Zeeman levels. At high magnetic fields the $J=2$
level shifts evolve nonlinearly with $H$ as a result of gap distortion 
by the magnetic field \cite{shi83}. For a discussion of these 
effects see \cite{sch83,fis86,hal90}.
Nevertheless, \cite{mov88} obtained a value of $g_{2^-}\simeq 0.04$
at $P=19\,\mbox{bar}$ from their measurements at fields above $1\,\mbox{kG}$,
in reasonable agreement with theoretical expectations.
However, as I show below the g-factor for the $J=2^{-}$ modes can be determined
with high acccuracy from an analysis of {\sl transverse} sound propagation
in a magnetic field. But, first I describe the mechanisms that lead to
a propagating transverse current mode.

\section{Transverse Modes}\label{tmodes}

The dispersion relation for a transverse current excitation
is given by the momentum conservation equation, Eq.
(\ref{Transverse_Current_Conservation_phi}),
and the response function for the stress induced by a 
transverse current fluctuation,
\be
\left(\frac{\omega}{v_fq}\right)_{\pm}=\twofifths
\left(1+F^s_1/3\right)\,\frac{1}{\sqrt{2}}
\left[\frac{\delta\phi_{2,\pm 1}}{\delta\phi_{1,\pm 1}}\right]_{\mbox{\tiny tot}}
\,,
\ee
where the subscripts $\pm$ refer to the two independent circular
polarizations of the transverse current. In addition to the direct
contribution to $\phi_{2,\pm 1}$ coming from the current fluctuations,
order parameter modes with $J=2$ and $M=\pm 1$ also 
induce transverse stress fluctuations and therefore couple to the 
transverse current. The result for the transverse stress is \cite{moo93},
\ber\label{phi-2pm}
\phi_{2,\pm 1} =
\onethird A^s_1\,\left(\frac{qv_f}{\omega}\right)\,
\xi_1(q,\omega)\,\frac{1}{\sqrt{2}}\,\phi_{1,\pm 1}
+
\left(\frac{\omega}{2\Delta}\right)\,\Lambda_1(q,\omega)\,
\cD^{(-)}_{2,\pm}
\,,
\eer
where the term $\sim A^s_1\,\phi_{1,\pm 1}$ represents the 
quasiparticle contribution to the transverse restoring force,
while the condensate contribution, which also enters through the 
Landau interaction, is proportional to the amplitude of the
$J=2^{-}$, $M=\pm 1$ order parameter modes, $\cD^{(-)}_{2,\pm}$.
I omit the $J=2^{+}$, $M=\pm 1$ modes
because the coupling of these modes is smaller by a factor of the 
particle-hole asymmetry ratio, $\zeta$.
Unlike the case of LZS, the $J=2^+$ modes
cannot be easily detected in the TZS spectrum because the TZS mode
is suppressed for frequencies below the $J=2^-$ mode, which lies above
the $J=2^+$ mode. The equations of motion for the $J=2^-$, $M=\pm 1$
order parameter modes are given by
\ber
\left[(\omega+i\Gamma)^2-\Omega_{2^-,\pm 1}^2-
\frac{2}{5}q^2v_f^2\right]\cD_{2,\pm}=
\left(\frac{8}{5}\Delta^2\right)\,
\left(\frac{qv_f}{2\Delta}\right)\,
\frac{A^s_1}{3}\,
\frac{1}{\sqrt{2}}\,
\phi_{1,\pm 1}
\,.
\eer
In zero field the dispersion relation for transverse current excitations
is independent of the polarization,
\be\label{TZS-dispersion}
\left(\frac{\omega}{qv_f}\right)^2=
\frac{F^s_1}{15}\,\rho_n(\omega)
+\frac{2 F^s_1}{75}\rho_s(\omega)\,
\left\{\frac{\omega^2}{(\omega+i\Gamma)^2 - 
\Omega_{2^-}^2 -\frac{2}{5}(q^2v_f^2)}\right\}
\,,
\ee
which displays contributions 
from the quasiparticle restoring force ($\sim\rho_n$),
and from the condensate ($\sim\rho_s$) \cite{mak77,moo93}.
Equation (\ref{TZS-dispersion})
is based on a long-wavelength expansion, i.e. $qv_f\ll \omega$, 
which is valid in the B-phase for frequencies in the range,
$\Gamma\ll |\omega-\Omega_{2^-}|\ll\omega$.\footnote{Results
for shorter wavelengths and frequencies far off resonance can be found
in \cite{moo93}.}

In normal \he~{transverse sound} propagates with a phase velocity 
$c_t\ge v_f$ provided the $\ell=1$ and $\ell=2$ Landau parameters satisfy,
$F^s_1/3 + F^s_2/(1+F^s_1/3) > 2$ \cite{fom68}; in the limit $F^s_1\gg 15$,
$c_t\simeq\sqrt{F^s_1/15}\,v_f$, which is the normal-state result in the 
limit $qv_f/\omega\ll 1$. The condition for a propagating TZS mode
is obeyed for pressures above approximately $1\,\mbox{bar}$;
however, the predicted velocity is close to the Fermi velocity at all pressures.
As a result it is difficult to differentiate transverse
oscillations of the Fermi surface from an incoherent current of quasiparticles;
and early attempts to observe transverse sound in the normal  phase of
$^3$He were inconlusive,~\cite{roa76,flo76}.

The onset of superfluidity in \he~has dramatic effects on the collisionless 
transverse modes in \he. The opening of a gap in the excitation
spectrum suppresses the quasiparticle contribution to the restoring
force. This suppression comes from the factor of $\rho_n$ in 
Eq. (\ref{TZS-dispersion}), which decreases rapidly below $T_c$.
For this reason early theoretical investigations emphasized that the 
transverse zero sound mode, even if observable in the normal state, would
disappear below $T_c$ \cite{mak74,com76a,mak77}.
The condensate also contributes to the dispersion relation through
the coupling to the $J=2^-$, $M=\pm 1$ order parameter modes. The 
condensate response suppresses TZS at low frequencies, as can 
be seen from Eq. (\ref{TZS-dispersion}). However, at intermediate
frequencies, $\Omega_{2^-} < \omega < 2\Delta$, the condensate provides
a strong restoring force through the Landau interaction energy, leading
to a propagating transverse current mode at low temperatures.
\begin{figure}[ht]
\centerline{\epsfxsize 1.2\hsize
\epsfbox{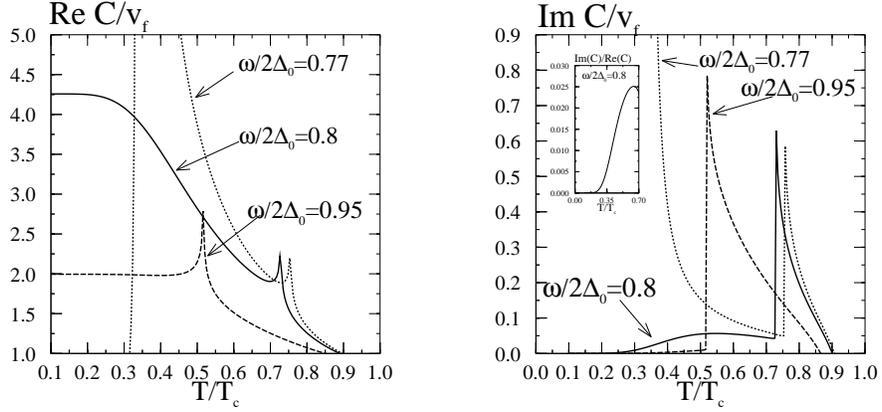}}
\caption{\small\label{TZS}
The temperature dependence of the phase velocity ($\mbox{Re}\,\cC/v_f$)
and attenuation ($\mbox{Im}\,\cC/v_f$) of the transverse current
mode in \heb~at the frequencies $\omega=1.9\Delta_0$, $\omega=1.6\Delta_0$ 
and $\omega=1.54\Delta_0$. A well defined propagating mode with low attenuation
exists at high frequency and low temperatures. At $\omega=1.54\Delta_0$ the TZS 
mode develops but is extinguished for $T<T_*\simeq 0.31 T_c$.
}
\end{figure}
The calculated temperature dependences of the phase velocity and
attenuation of a transverse current excitation in \heb~are shown in
Fig. \ref{TZS} taken from Ref. \cite{moo93}.
At low frequency, $\omega\ll\Omega_{2^-}(0)$, 
the transverse current mode is highly attenuated
over the full temperature range below $T_c$.
At high frequencies, $\omega = 1.9\Delta_0 > \Omega_{2^-}(0)$, a propagating
transverse current mode develops with a large phase velocity and low 
absorption that decreases exponentially at low temperature. For intermediate
frequency, $\omega=1.54\Delta_0$, TZS develops for $T<T_{\mbox{\tiny pb}}$;
the attenuation drops until $T=T_*$, at which point TZS is extinguished by
resonant absorption from the $J=2^-$ modes.

\vspace*{12pt}
\noindent{\large\it Circular Birefringence of Transverse Waves}
\vspace*{12pt}

The B-phase of \he~is symmetric under time-inversion, and consequently is
non-birefringent for transverse wave propagation, i.e. 
right- and left-circularly polarized waves propagate at the same
velocity. However, {\it circular birefringence}, i.e. $\cC_+\ne\cC_-$,
can be induced by a magnetic field. If the field is applied parallel to the
direction of propagation of the waves, $\vH ||\vq$, then axial symmetry
is preserved and the eigenmodes for the transverse current are the 
right- and left-circular polarization states. However, the eigenfrequencies
of these modes, or equivalently the complex phase velocities,
are no longer degenerate. This leads to the acoustic analog of 
(circular) optical birefringence ($\mbox{Re}\,\cC_+\ne \mbox{Re}\,\cC_-$) and 
dichroism ($\mbox{Im}\,\cC_+\ne \mbox{Im}\,\cC_-$) \cite{moo93}.
At low temperatures and high frequencies, $\omega>\Omega_{2^-,\pm 1}$,
circular birefringence gives rise to the {\sl acoustic Faraday effect} in 
which the direction of a linearly polarized transverse wave rotates along 
the direction of propagation.

The dispersion relations for RCP $(+)$ and LCP $(-)$ transverse current
modes in a magnetic field, $\vH ||\vq$, are given by
\be
\left(\frac{\omega}{qv_f}\right)_{\pm}^2 =
\Lambda_n + \Lambda_s
\left(
\frac{\omega^2}{(\omega+i\Gamma)^2-\Omega^2_{2^-,\pm 1}(T,\omega,H)}
\right)
\ee
where $\Lambda_n=(F^s_1/15)\,\rho_n(\omega,T)$ and
$\Lambda_s=(2F^s_1/75)\,\rho_s(\omega,T)$ in the long-wavelength limit.
The condensate term dominates at low temperature ($\Lambda_s\gg\Lambda_n$)
and is anomalously large when the sound frequency is nearly resonant with the
$J=2^-$, $M=\pm 1$ modes,
\be
D^2_{\pm}(\omega,H,T)=(\omega + i\Gamma)^2 - 
\Omega^2_{2^-,\pm 1}(\omega,H,T)\,\approx\,0
\,.
\ee
The frequencies of the $M=\pm 1$ modes include the Zeeman shifts,
\be
\Omega^2_{2^-,\pm 1}=
\Omega_{2^-}^2(T)\pm 2\,g_{2^-}(T)\,\omega\,\omega_{\mbox{\tiny eff}}(H,T)
\,,
\ee
where $\Omega_{2^-}(T)$ is the frequency of the $J=2^-$ modes in zero field.

Consider a transverse current excitation of frequency $\omega$
at $z=0$, linearly polarized along $\vx$ and propagating in the $z$-direction,
$\vJ(\omega,z=0) = \cJ(\omega)\,\hat{\vx}$.
The RCP and LCP modes propagate with different phase velocities; thus
the current evolves as,
\be
\vJ(\omega,z) = 
\frac{\cJ}{\sqrt{2}}\,\,e^{iq_+(\omega)z}\,\hat{\ve}_{+}
\,+\,
\frac{\cJ}{\sqrt{2}}\,\,e^{iq_-(\omega)z}\,\hat{\ve}_{-}
\,,
\ee
where $\hat{\ve}_{\pm}=(\hat{\vx}\pm i\hat{\vy})/\sqrt{2}$
are the polarization vectors for RCP and LCP current modes.
This response corresponds to a pure Faraday rotation of the polarization
if the phase velocities are real. This is the case
for low temperatures, $T\ll T_c$, and frequencies above the
collective mode resonances, $\omega > \Omega_{2^-,\pm 1}(T)$.
The spatial period of the rotation of the polarization 
is identified by writing
\be\label{Faraday_Rotation_J}
\vJ(\omega,z) = 
\cJ\,e^{iq(\omega)z}\, 
\left( \cos(\delta q z)\hat{\vx}
\,-\,
\sin(\delta q z)\hat{\vy} \right)
\,,
\ee
where the sum and difference wavevectors are
$q=(q_{+} + q_{-})/2$ and $\delta q=(q_{+} - q_{-})/2$.
The linearly polarized transverse wave propagates 
with a mean phase velocity, $\bar{c}_t = \omega/q$,
and the polarization {\em rotates} with the spatial period,
\be
\Lambda(\omega,H,T) = \frac{2\pi}{|\delta q|}
                    = \frac{4\pi}{q_{-} - q_{+}} 
\,.
\ee

Near the $M=+1$ mode crossing the wavenumber for the RCP
wave vanishes; $q_+(T_+)=0$, since $D_{+}(\omega,T_{+}) = 0$ and
$\Lambda_n\ll\Lambda_s$.
The temperature dependence of the Faraday
rotation period is then determined by the response of this mode {\it off
resonance},
\be
q_{+}(T)\simeq q_f\,\frac{D_{+}(T)}{\sqrt{\Lambda_s}\omega} = 
               q_f\,\sqrt{s(T/T_{+} -1)}
\,,
\ee
where $q_f=\omega/v_f$, 
\be
s\equiv\frac{2|\Omega_{+}^{'}|T_{+}}{\Lambda_s\Omega_{+}}
\,,
\ee
$\Omega_{+}=\Omega_{2^-}(T_+)=\omega$, and $\Omega_{+}'=d\Omega_{2^-}(T)/dT|_{T_+}$
are evaluated at the mode resonance, $T=T_{+}$. The field dependence originates
from the linear Zeeman splitting of the $M=\pm 1$ modes, and enters via the
wavenumber of the $M=-1$ mode,
\be
q_{-}\simeq q_f\sqrt{ \frac{2|\Omega_{+}^{'}|(T-T_{+})+
4g_{2^-}\omega_{\mbox{\tiny eff}}}{\Lambda_s\Omega_{+}}}
\,,
\ee
for $T_+\lsim T\ll T_c$.
Thus, the Faraday rotation period varies with field on the scale,
\be
B_{+}\equiv \frac{\Lambda_s\Omega_{+}}{4\,\gamma_{\mbox{\tiny eff}}\,g_{2^-}}
\,,
\ee
where $\gamma_{\mbox{\tiny eff}}\equiv\omega_{\mbox{\tiny eff}}/H$
is the effective gyromagnetic ratio at the mode crossing.
I can express the temperature and field dependence of $\Lambda$
near the $J=2^-$, $M=+1$ mode crossing in terms of scaled temperature
and field variables, $\alpha = s\,(T/T_{+} -1)$ and $\beta = H/B_{+}$,
\be\label{Faraday_Period_Theory}
\frac{\Lambda}{\lambda_f}=\frac{2}{\sqrt{\alpha+\beta}-\sqrt{\alpha}}
=
\Bigg\{\matrix{4\sqrt{\alpha}/\beta &,\quad\beta\ll\alpha \cr
     2\left(1+\sqrt{\alpha/\beta}\right)/\sqrt{\beta}
                                             &,\quad\beta\gg\alpha\,,}
\ee
where $\lambda_f=2\pi v_f/\Omega_{+}$.
For $T_+/T_c\simeq 0.4$, $\omega/2\pi=80\,\mbox{MHz}$ and $F^s_1\simeq 15$
the field scale is $B_+\simeq 1\,\mbox{Tesla}$.
Thus, except for a narrow temperature range
near resonance, i.e. $|T-T_+| < 10^{-2}T_+$ at $H=100\,\mbox{G}$,
the Faraday rotation period varies as
\be\label{Verdet}
\Lambda 
\simeq 4\,\frac{B_+}{H}\,\sqrt{s(T-T_+)}
\simeq \frac{\xi_{\Omega_+}\pi v_f}{\gamma_{\mbox{\tiny eff}} g_{2^-}}
\sqrt{\frac{2}{75}F^s_1\rho_s}\,\frac{\sqrt{T-T_+}}{H}
\,,
\ee
with $\xi_{\Omega}=8|\Omega_{+}'|T_{+}/\Omega_{+}\simeq 1$ at $T\simeq 0.44 T_c$.
In magneto-optics the field-independent prefactor,
$V=\Lambda H/2\pi$, is called the Verdet constant.

\vspace*{12pt}
\noindent{\large\it Broken Spin-Orbit Symmetry}
\vspace*{12pt}

The acoustic Faraday effect provides a clear
demonstration of {\sl spontaneously broken spin-orbit symmetry}.
The direction of mass flow is altered by the application of a magnetic field
along the propagation direction.
The role of broken spin-orbit symmetry is evident in the
the contribution to the stress fluctuations from the order parameter,
\be
\delta\pi_{ij}\sim \sum_M\,\cD^{(-)}_{2,M}\,\vt_{ij}^{2,M}
\,.
\ee
Inspection of Eqs. (\ref{Spherical_Tensors}) shows that the transverse 
stress couples the transverse current fluctuations only to the $M=\pm 1$
modes with {\sl total} angular momentum $J=2$,
\be
\vt_{ij}^{(2,\pm 1)}\sim (\hat\vq_i\ve^{\pm}_j + \ve^{\pm}_i\hat\vq_j)
\,.
\ee
Broken time-reversal symmetry is also important for producing circular
birefringence. The Zeeman splitting of the $M=\pm 1$ modes in a field 
aligned along $\vq$ produces a difference in the stress induced by the
$M=+1$ and $M=-1$ modes, and thus, different propagation velocities.

These general symmetry considerations also imply that superfluid \hea, which
spontaneously breaks time-reversal symmetry as well as relative spin-orbit
rotation symmetry, should exhibit spontaneous acoustic circular
birefringence or dichroism for transverse excitations propagating along 
the {\boldmath{$\ell$}} vector. In fact observation of this effect for a uniform
texture would provide a direct evidence of broken time-reversal symmetry in
\hea~\cite{yip92}.

\section{Observation of Transverse Current Waves in \heb}

Propagating transverse current waves were recently observed in 
\heb~at low pressure \cite{lee99}.
The excitation and detection of these waves was made in 
an acoustic cell with a transducer operating at a 
frequency of $\omega/2\pi=82.26\,\mbox{MHz}$ that {\sl generates and
detects} transverse currents with a fixed linear polarization.
In a high attenuation regime the acoustic response measures the 
transverse acoustic impedance of bulk \he. However, if a propagating
mode develops with low attenuation then current waves propagate 
across the cell, are reflected from the boundary opposite the
transducer and are detected by the same transducer. In this regime
standing waves develop and are observable in the acoustic impedance
because of interference between the source and 
reflected wave at the detector. Oscillations of the impedance
occur as the phase velocity and attenuation change with temperature, pressure
or field. A linearly polarized wave with $\vq\parallel\vH$ undergoes Faraday 
rotation of its polarization as it propagates.
Upon reflection from the opposite wall of the cavity
the reflected wave propagates with its polarization 
rotating with the same handedness
relative to the direction of the field, i.e. the rotation of the polarization
accumulates after reflection from a surface. The rotation of
the polarization produces an oscillatory modulation of the impedance
as a function of magnetic field, with the period of the impedance oscillations
being determined by, $\Lambda/2\propto 1/H$.
\begin{figure}[ht]
\centerline{\epsfxsize= 0.8\hsize
\epsfbox{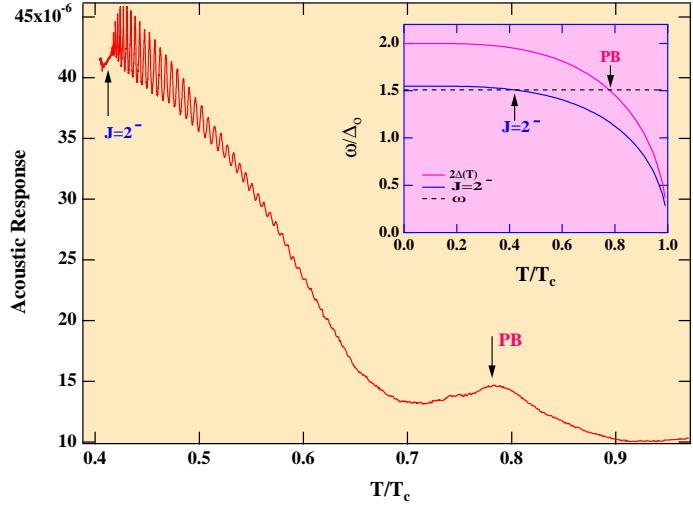}}
\caption{\small\label{transverse_impedance}
Transverse impedance oscillations in \heb~at $\omega/2\pi=82.26\,\mbox{MHz}$,
$P=4.31\,\mbox{bar}$ and zero field.
The oscillations develop in the intermediate frequency range
$\Omega_{2^-}<\omega<2\Delta(T)$, between the pairbreaking edge, labeled
by PB ($\omega=2\Delta(T_{\mbox{\tiny PB}})$), and
the extinction point determined by the resonance frequency of the $J=2^-$,
$M=\pm 1$ modes, i.e. $\omega=\Omega_{2^-}(T_*)$.
The pair-breaking and resonance conditions 
for fixed frequency are shown in the inset.
}
\end{figure}

The transverse acoustic response measured by \cite{lee99}
is shown for \he~at a pressure of 
$4.31\,\mbox{bar}$ in Fig. \ref{transverse_impedance}. The
impedance varies smoothly as the temperature drops below $T_c$ and shows a
peak at the pairbreaking (PB) edge, $\omega=2\Delta(T_{\mbox{\tiny pb}})$.
Oscillations develop below a temperature of $T\simeq 0.65T_c$,
indicating the presence of a propagating mode.
The amplitude of the impedance oscillations increases as the 
temperature decreases indicating that the attenuation of the
propagating mode decreases with decreasing temperature.
The oscillations disappear at the temperature $T_*\simeq 0.42\,T_c$
corresponding to the resonance condition,
$\omega=\Omega_{2^-}(T_*)$; i.e. transverse waves are extinguished below
this temperature. All of these features are in agreement with
theoretical predictions \cite{moo93}.
\begin{figure}[ht]
\centerline{\epsfxsize=1.2\hsize
\epsfbox{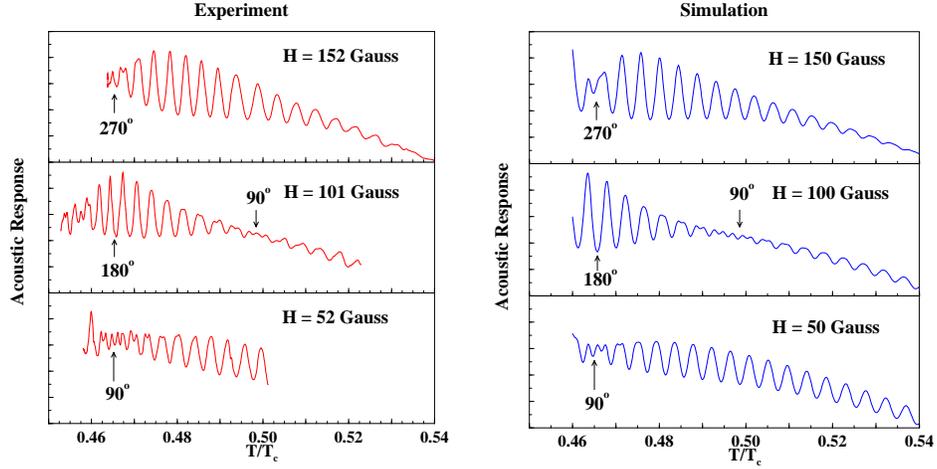}}
\caption{\small\label{Faraday_Rotation_Data}
Left panels: Magnetic field dependence of the acoustic cavity response 
measured at 4.42~bar. The angles indicate
the rotation of the polarization after one round-trip.
Right panels: Simulation of the acoustic impedance for the same
path length, temperature, pressure and magnetic fields.
}
\end{figure}

Figure \ref{Faraday_Rotation_Data} (reproduced from \cite{lee99})
shows the impedance oscillations at a pressure of $4.42\,\rm{bar}$
for magnetic fields of $52\,\rm{G}$, $101\,\rm{G}$ and $152\,\rm{G}$.
The magnetic field modulates the zero field impedance oscillations.
The transducer detects only transverse
waves having the same linear polarization as the source wave. A field of
$52\,\rm{G}$ suppresses the impedance oscillations near $T=0.465\,T_c$
that were present in zero field, corresponding to a
$90^{\circ}$ rotation of the polarization of the reflected
wave; i.e. the polarization of the reflected wave is orthogonal to 
the detection direction. Doubling the field restores the impedance 
oscillations near $T=0.465\,T_c$. The oscillations are suppressed again
at a field to $152\,\rm{G}$. Near the $90^{\circ}$ and $270^{\circ}$ points, 
there are small amplitude impedance oscillations, with shorter period than the
primary oscillations. These oscillations are due to interference 
of the source and waves that traverse the cavity twice. This interpretation
was verified by comparing the impedance oscillations with the 
calculated cavity response, shown in the right panels of 
Fig. \ref{Faraday_Rotation_Data}.
The calculation is based on Eqs. (\ref{Faraday_Rotation_J}) and (\ref{Verdet}) and 
takes as input the measured attenuation and phase velocity in zero field.
The Verdet constant is obtained from the measurement at $52\,\mbox{G}$.
The simulation reproduces the observed features of the
impedance as a function of temperature including
the maximum in the modulation at $T/T_c=0.465\,,\,H=101\,\mbox{G}$
and the minimum at $T/T_c=0.465\,,\,H=152\,\mbox{G}$,
as well as the fine structure oscillations in the impedance
near the points labeled $90^{\circ}$ and $270^{\circ}$.
The fine structure is observed when the polarization rotates by an
odd multiple of $90^{\circ}$ upon a single round trip in the cell.
Waves that traverse the cell
twice are then $180^{\circ}$ out of phase relative to the source wave,
and consequently the period of the impedance oscillations is halved.
The amplitude of the oscillations is reduced
because of attenuation over the longer pathlength.
These observations of the rotation of the polarization of the current
by a magnetic field prove that the impedance oscillations result from
the interference of propagating {\sl transverse} current waves \cite{lee99}.

\vspace*{12pt}
\noindent{\it The g-factor for the $J=2^-$ Modes}
\vspace*{12pt}

\begin{figure}[ht]
\centerline{\epsfxsize=0.7\hsize
\epsfbox{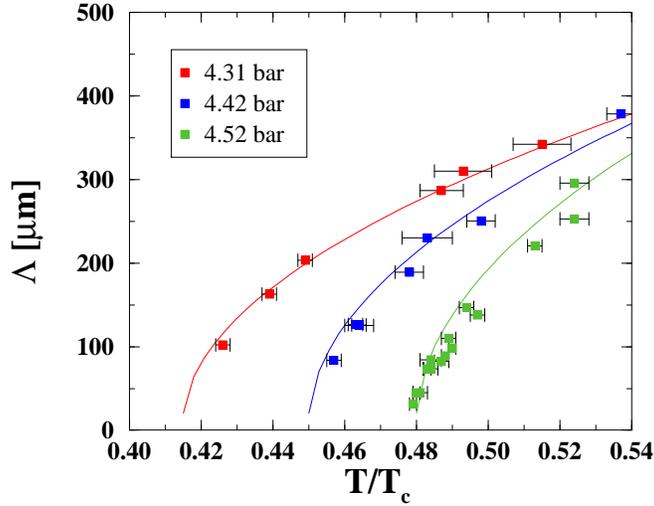}}
\caption{\small \label{Faraday_Lambda}
Comparison between the theoretical and experimental results for the 
temperature dependence of the Faraday rotation period for a field
of $H=100\,\mbox{Gauss}$ at pressures, $p=4.31\,\mbox{bar}$ (red),
$p=4.42\,\mbox{bar}$ (blue), and $p=4.52\,\mbox{bar}$ (green). The data for
the corresponding pressures are shown as the solid squares with error
bars (Y. Lee, et al. (1999)).}
\end{figure}

The temperature and field variations of the impedance oscillations provide
a strong test of the theoretical predictions for transverse wave propagation
in \heb. Here I compare the theoretical result for $\Lambda$ in 
Eq. (\ref{Faraday_Period_Theory}) 
with the experimental data for magneto-acoustic rotation
period for low fields, $H \ll 1\,\mbox{kG}$, and temperatures 
above, but near, the extinction point $T_*$.
Note that the temperature, $T_{+}$, is the extinction
point in a field is slightly higher than the 
zero-field extinction point, $T_*$; 
at $H=100\,\mbox{G}$, $T_{+}-T_*\approx 1~{\mu}K$.
The magnitude of the Faraday rotation period depends on well known 
properties of \heb, except for the Land\'e g-factor, $g_{2^-}$. For the calculations
shown in Fig. \ref{Faraday_Lambda} I used the measured Fermi-liquid data
tabulated for $T_c$, $v_f$, $F_1^s$, $F_0^a$, and $F_2^a$ as a function of pressure 
in \cite{hal90}. The mode data are $T_{+}/T_c$ and $\Omega_{+}(T_{+})=\omega$, and 
$\gamma=3.2434\,\mbox{MHz/kG}$. The only undetermined parameter that
enters Eq. (\ref{Faraday_Period_Theory}) for $\Lambda$ is the Land\'e g-factor,
which scales the magnitude of the Faraday rotation angle. The data exhibit
the temperature dependence and pressure dependence predicted by the theory and
provide a new determination of the Land\'e g-factor the 
$J=2^-$ modes at low fields. The calculated results
are in excellent agreement with the experimental data at three nearby pressures
for $g_{2^-}=0.02\pm 0.002$. 

This result differs from the value of $g_{2^-}\simeq 0.042$ 
reported by \cite{mov88} for $p=19\,\mbox{bar}$, which was
obtained from the splitting of the $J=2^-$
multiplet in the absorption spectrum of longitudinal sound
for fields above $2\,\mbox{kG}$. At these high fields
the non-linear field contribution to the level splittings
is comparable to the linear Zeeman splitting,
which makes and accurate determination of the Land\'e g-factor from the
absorption spectrum difficult. This complication is not present in the
analysis of the Faraday rotation period at $H\approx 100\,\mbox{G}$.

\begin{figure}[ht]
\centerline{\epsfxsize=0.6\hsize
\epsfbox{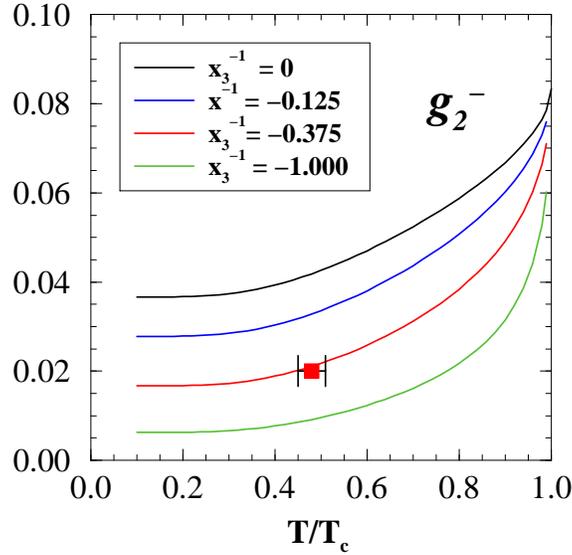}}
\caption{\small\label{g2-}
Comparison between the theoretical and experimental results for the 
Land\'e g-factor of the $J=2^-$ modes. The theoretical curves are
include the effects of f-wave pairing correlations. The experimental result
for $g_{2^-}$ at $T/T_c=0.46$ and $P=4.32\,\mbox{bar}$ is fit
with an attractive f-wave interaction, 
$x_3^{-1}\simeq -0.33$.}
\end{figure}

Earlier theoretical calculations for the order parameter collective mode
frequencies showed that that the g-factors for these excitations are 
sensitive to Fermi liquid interaction effects and higher angular momentum
pairing correlations \cite{sau82}. Figure \ref{g2-} shows the
g-factor for the $J=2^-$ modes as a function of temperature
for several values of the f-wave pairing interaction, expressed in terms
of the f-wave and p-wave transition temperatures,
$x_3=\ln(T_{\mbox{\tiny f-wave}}/T_c)$. 
Negative values correspond an attractive f-wave pairng interaction.  
The g-factor, $g_{2^-}$, also depends on the Landau interaction parameter,
$F_2^s$, which is known from measurements of the normal-state zero sound velocity.
The calculated values of $g_{2^-}$ are shown for a pressure of 
$4.31\,\mbox{bar}$, corresponding to $F^s_2\simeq -0.05$. 
The calculated g-factor is strongly reduced by attractive f-wave
correlations. 

The analysis of the Faraday rotation period yields a measurement
of $g_{2^-}$. The fact that the
$J=2^-$ g-factor is intrinsically small (in contrast to the $J=2^+$
g-factor), makes this parameter a sensitive  test of many-body
correlation effects in superfluid \he. The implications of this result
are that either substantial attractive f-wave pairing correlations
or strong-coupling effects reduce the g-factor by a factor of two
compared to the weak-coupling result.
The value of $g_{2^-}$ obtained from Faraday rotation measurements suggests 
that f-wave pairing correlations at $4\,\mbox{bar}$ are attractive with
$T_{\mbox{\tiny f-wave}}\simeq 0.07\,T_c$.
If the reduction is due to
f-wave pairing correlations then there should be a
collective mode with total angular momentum $J=4$ \cite{sau81a} 
lying just below the pair breaking edge which might be observable in
the dispersion of TZS at high frequencies near, but below the pair breaking edge.
In summary, the discovery of propagating transverse currents and magneto-acoustic
rotation of the polarization of mass currents in superfluid \heb~opens
a new spectroscopy of collective excitations in superfliud \heb. 
The precision of this spectroscopy offers the possibility of measuring
strong coupling effects on the collective mode frequencies and Zeeman levels,
as well as effects of higher angular momentum pairing correlations at 
ultra-low temperatures.

\vspace*{12pt}
\noindent{\it Acknowledgements}
\vspace*{12pt}

I thank my colleagues and collaborators, Geneva Moores, Yoonseok Lee, Tom Haard,
and Bill Halperin for their many contributions included in these lecture notes,
and for many important discussions on acoustics and superfluid \he.

\bibliography{master}
\bibliographystyle{klunamed}

\end{document}